\documentclass[prd,twocolumn,nofootinbib]{revtex4}
\usepackage{graphicx, epsfig}
\usepackage{color}
\usepackage{mathrsfs}
\usepackage {amssymb}
\usepackage {amsmath}

\newcommand{\bq}{\begin{eqnarray}}
\newcommand{\eq}{\end{eqnarray}}
\newcommand{\bsq}{\begin{subequations}}
\newcommand{\esq}{\end{subequations}}
\newcommand{\bc}{\begin{center}}
\newcommand{\ec}{\end{center}}

\newcommand{\nc}{\newcommand}
\newcommand{\nn}{\nonumber}
\nc{\ba}{\begin{eqnarray}}
\nc{\ea}{\end{eqnarray}}
\newcommand\be{\begin{equation}}
\newcommand\beq{\begin{equation}}
\newcommand\ee{\end{equation}}
\newcommand\eeq{\end{equation}}

\nc{\ga}{\gamma}
\nc{\tnu}{\bar{\nu}}
\nc{\tmu}{\bar{\mu}}
\nc{\tq}{\tilde{q}}

\nc{\x}{{\bf{x}}}
\nc{\bmi}{\bar \mu_{i} }
\nc{\bm}{\bar \mu }
\newcommand{\gsim}{\raise.3ex\hbox{$>$\kern-.75em\lower1ex\hbox{$\sim$}}}
\newcommand{\lsim}{\raise.3ex\hbox{$<$\kern-.75em\lower1ex\hbox{$\sim$}}}
\nc{\gm}{\gamma }
\nc{\PIJ}{P_{,X_{IJ}}}
\nc{\hP}{\hat{P}}
\nc{\PX}{P_{,X}}
\nc{\half}{\frac{1}{2}}
\nc{\p}{\phi}

\nc{\s}{\sigma}

\nc\bfphi{{\bf \phi}}

\input{epsf.sty}

\begin{document}

%%%%%%%%%%%%%%%%%%%%%%%%%%%%%%%%%%%%%%%%%%%%%%%%%%%%%%%%%%%%%%%%%%%%%%%

\title{Constraints on the fundamental string coupling from B-mode experiments}

\author{A.~Avgoustidis$^1$, E.~J.~Copeland$^2$,  A.~Moss$^3$, L.~Pogosian$^4$, A.~Pourtsidou$^5$,  and Dani\`ele A.~Steer$^{6}$}

\affiliation{$^1$Centre for Theoretical Cosmology, DAMTP, CMS, Wilberforce Road, Cambridge CB3 0WA, UK}
\affiliation{$^2$School of Physics and Astronomy, University of Nottingham, University Park, Nottingham NG7 2RD, UK}
\affiliation{$^3$ Department of Physics and Astronomy, University of British Columbia, Vancouver, BC, Canada}
\affiliation{$^4$Department of Physics, Simon Fraser University, Burnaby, BC, V5A 1S6, Canada}
\affiliation{$^5$Jodrell Bank Centre for Astrophysics, 
%School of Physics and Astronomy, 
University of Manchester, Manchester M13 9PL, UK}
\affiliation{$^6$APC, UMR 7164, 10 rue Alice Domon et L\'eonie Duquet,75205 Paris Cedex 13, France}

\date{\today}

\begin{abstract}
We study signatures of cosmic superstring networks containing strings of multiple tensions and Y-junctions, on the cosmic microwave background (CMB) temperature and polarisation spectra. Focusing on the crucial role of the string coupling constant $g_s$, we show that the number density and energy density of the scaling network are dominated by different types of string in the $g_s \sim 1$ and $g_s \ll 1$ limits. This can lead to an observable shift in the position of the B-mode peak --- a distinct signal leading to a direct constraint on $g_s$. We forecast the joint bounds on
$g_s$ and the fundamental string tension $\mu_F$ from upcoming and future CMB polarisation experiments, as well as the signal to noise in detecting the difference between B-mode signals in the limiting cases of large and small $g_s$. We show that such a detectable shift is within reach of planned experiments. 
\end{abstract}

\maketitle

The extraordinary wealth of cosmological data of increasing precision is allowing us to seriously investigate models of the very early universe. Here we focus on aspects of one promising model, namely the potential cosmic microwave background (CMB) signatures of cosmic superstrings -- small tension  line-like objects stretched over cosmological scales -- whose existence is characteristic of a class of inflationary models based on string theory.  The observational properties of these strings depend sensitively on the value of the string coupling $g_s$, a fundamental parameter of string theory, and here we show for the first time that it may be possible to constrain this parameter through future CMB polarisation data.

Cosmic superstrings differ from their topological counterparts, i.e. field theory cosmic strings \cite{Kibble}, in a number of ways. First, they have a spectrum of tensions $\mu_i$ determined by $g_s$ and the  $(p_i,q_i)$ charges they carry \cite{Schwarz:1995dk,Witten:1995im}:
$\mu_i=(\mu_F/g_s) \sqrt{p_i^2 g_s^2+q_i^2}$, where $\mu_F$ is the tension of the lightest (1,0) fundamental string. Second, their intercommutation probabilities can be significantly less than unity \cite{Jones:2003da, Jackson:2004zg, Hanany:2005bc}, so when two cosmic superstrings collide they do not necessarily reconnect.  Third, when strings of {\it different} charges collide, Y-junctions -- i.e. vertices joining 3 string segments 
of different charge -- can form \cite{CopKibSteer1}.
In order to generalise previous work on the B-mode CMB polarisation 
signal from standard cosmic strings \cite{Pogosian:2007gi} to cosmic superstrings, a first step is therefore to solve for the dynamics of a multi-tension string network with junctions.

To do so we use the network evolution model developed in \cite{Avgoustidis:2009ke,NAVOS} in which the fundamental variables are $\rho_i$ and $v_i$, respectively the energy density and root-mean-squared (rms) velocity of strings of type $i$, with correlation length defined by $\rho_i=\mu_i L_i^{-2}$. They obey the coupled equations 
\be\label{v_idtgen} 
\dot v_i = (1-v_i^2)\left[\frac{1}{L_i}\left(\frac{2\sqrt{2}(1-8v_i^6)}{\pi(1+8v_i^6)}\right)-2\frac{\dot a}{a}v_i \right] \,  
\ee
\ba
    \dot\rho_i &=& -2\frac{\dot a}{a}(1+v_i^2)\rho_i-\frac{c_i 
    v_i\rho_i}{L_i} 
    \nn \\ 
    && - \sum_{a,k} \frac{d_{ia}^k   
    \bar v_{ia} \mu_i \ell_{ia}^k(t)}{L_a^2 L_i^2} + \sum_{b,\,a\le b}   
    \frac{d_{ab}^i \bar v_{ab} \mu_i   
    \ell_{ab}^i(t)}{L_a^2 L_b^2}      
    \label{rho_idtgen}    \ ,
\ea
where $\cdot = d/dt$ and $a(t)$ is the Friedmann-Lema\^itre-Robertson-Walker (FRW) universe scale factor.

The presence of Y-junctions is reflected in the last two terms in (\ref{rho_idtgen}); the penultimate one, for example, models the energy loss from the strings of type $i$ due to their collision with strings of type $a$ (with average relative velocity $\bar{v}_{ia}$), resulting in the formation of type $k$ links of average length $\ell_{ia}^k(t)$.  This occurs with $g_s$-dependent probability $d_{ia}^k=d_{ai}^k$ which depends both on the microphysical quantum intercommutation probability ${\cal P}_{ia}$ calculated in \cite{Jackson:2004zg,Hanany:2005bc}, and on a purely classical kinematic constraint $S_{ia}^k$ described in \cite{Avgoustidis:2009ke,CopKibSteer1}.
Self-interactions of strings of type $i$, leading to the formation of loops, are quantified by the coefficients $c_i$ in (\ref{rho_idtgen}). These are determined by the microphysical self-intercommutation probabilities ${\cal P}_{ii}$ and are also $g_s$-dependent.
Finally, all the ${\cal P}_{ij}$ also incorporate a model-dependent volume factor described by another parameter $w$~\cite{cmbmulti}, which in this Letter we set to unity.

Following the results~\cite{IntProbs} of simulations of single-string networks with varying intercommuting probabilities (and no junctions $d_{ij}^k=0$), we have taken $c_i =\tilde c\times {\cal P}_{ii}^{1/3}$. Matching to single-string network simulations with ${\cal P}_{ii}=1$,  then fixes $\tilde{c}=0.23 \; (0.18)$ in the radiation (matter) era \cite{VOSk}.  
There are at present no network simulations with which to calibrate interactions between strings of different types ($i\ne j$) , but one expects a similar dependence on ${\cal P}_{ij}$ so we take $d_{ij}^k = S_{ij}^k \times {\cal P}_{ij}^{1/3}$ \cite{cmbmulti}.

\begin{figure}[tbp]
\begin{center}
\includegraphics[height=8cm, width=7.7cm]{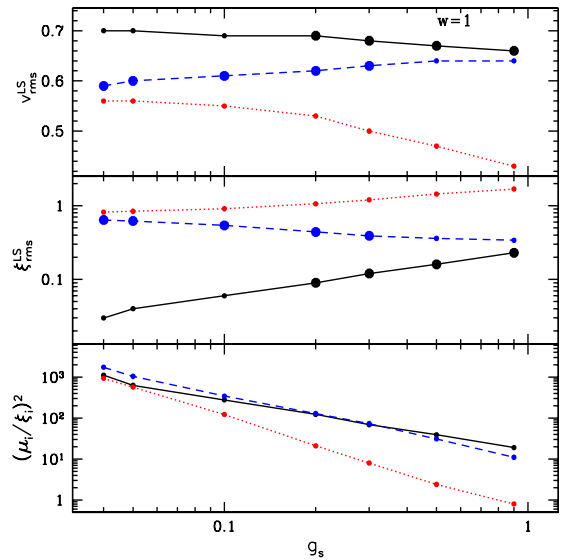}
\caption{\label{figNetwork} Dependence of the rms velocity (top), correlation length (middle) and power spectrum density (bottom panel) at the time of last scattering (LS) on the value of the string coupling $g_s$ for the F strings (solid black lines), D strings (blue dashed lines) and FD strings (red dotted lines). The rms velocity and correlation length of the string type(s) dominating the power spectrum are shown with oversized dots.}
\end{center}
\end{figure}

Despite the infinite hierarchy of Eqs.~(\ref{v_idtgen}-\ref{rho_idtgen}), in practice only the first few lightest strings dominate \cite{NAVOS,Avgoustidis:2009ke,TWW}, so we truncate the system at $i=7$ keeping all coprime charge pairs up to $(1,3)$ and $(3,1)$. 
The generic attractor solution of this system, with ${\cal P}_{ij}$ and $S_{ij}^k$ as computed in \cite{cmbmulti}, is a \emph{scaling} one in which  $\xi_i\equiv L_i/t$ and $v_i$ asymptotically approach constant values.  For all studied values of the string coupling $g_s$, the three lightest strings, namely $(1,0)$ F strings, $(0,1)$ D strings and $(1,1)$ FD strings, dominate the string number and energy density of the network. This is due to the interactions between the strings, and is expected from microphysical studies of junction dynamics \cite{CopKibSteer1}.
Interestingly, however, while the hierarchy in the {\it string number density}
$N_i = \xi_i^{-2}$ 
remains the same (i.e.~F strings are the most populous for all $g_s$, followed by D and FD strings), the {\it power spectrum density} \cite{cmbmulti} of the network $M_i\equiv \left({\mu_i}/{\xi_i}\right)^2$
becomes dominated by the heavy D strings at small string couplings, despite the fact that they are rare. In Fig.~\ref{figNetwork} we plot the rms velocity, the correlation length and the power spectrum density at the time of last scattering (LS) as a function of the string coupling $g_s$ for the three lightest network components. The power spectrum density is dominated by the light populous F strings for $g_s \lesssim 1$ and by the heavy rare D strings for $g_s \ll 1$. This general trend appears to be quite robust \cite{cmbmulti}, and is one of the main reasons behind the observable change in the B-mode spectrum. The other key ingredient is the fact that the correlation length of the D strings increases with decreasing $g_s$, as shown in Fig.~\ref{figNetwork}. This increase is greater for the weaker dependence of $d_{ij}^k$ coefficients on ${\cal P}_{ij}$. Further investigation is needed to confirm the functional dependence of the $d_{ij}^k$ coefficients on ${\cal P}_{ij}$, and to understand better the non-perturbative interactions that determine ${\cal P}_{ij}$ for heavy strings -- the two main sources of uncertainty in our modelling.

With the scaling solutions for interacting F-D networks at hand, we can predict their temperature (TT) and B-mode polarisation (BB) spectra. This is done using a generalised version of CMBACT \cite{Pogosian:1999np} for multi-tension string networks modelled by equations~(\ref{v_idtgen}-\ref{rho_idtgen}), and imposing that the total TT contribution from strings cannot exceed $10\%$ of the observed CMB anisotropy \cite{Wyman:2005tu,Bevis:2007gh,Battye:2010xz}. The amplitude of the CMB angular spectra $C_{\ell}$ for these multi-tension networks can be written as the sum of the power spectrum densities $M_i$ of each string type: 
\be
C_\ell^{strings} \propto M_{\rm total} = \sum_{i=1}^{N} M_i = \sum_{i=1}^{N} \left(\frac{\mu_i}{\xi_i}\right)^2 ,
\label{clstringN}
\ee
so this ultimately constrains the fundamental string tension $\mu_F$.
Here, we adjust $\mu_F$ to be such that
\be\label{fs}
f_s=C^{TT}_{strings} / C^{TT}_{total} < 0.1 \,,
\ee
where $C^{TT}_{strings}$ and $C^{TT}_{total}$ are defined by summing 
the corresponding $C_\ell$'s over multipoles $2\le\ell\le2000$ \cite{Pogosian:2007gi}.

Because of the important $\emph{vector}$ mode contribution from strings, the string induced B-mode signal can be much stronger than the one from inflationary tensor modes \cite{Seljak:1997ii, Pogosian:2007gi} and, indeed, strings can be a prominent source of B-mode polarisation on subdegree scales even with a marginal TT contribution of $1\%$ \cite{cmbmulti}.  The shapes of the spectra depend on the large-scale properties of the network, namely the correlation length and the rms velocity. In particular, the position of the main B-mode peak moves towards lower $\ell$ for larger correlation length, while the dependence on the rms velocity is nonlinear.

\begin{figure}[h]
\centering
\includegraphics[scale=0.5]{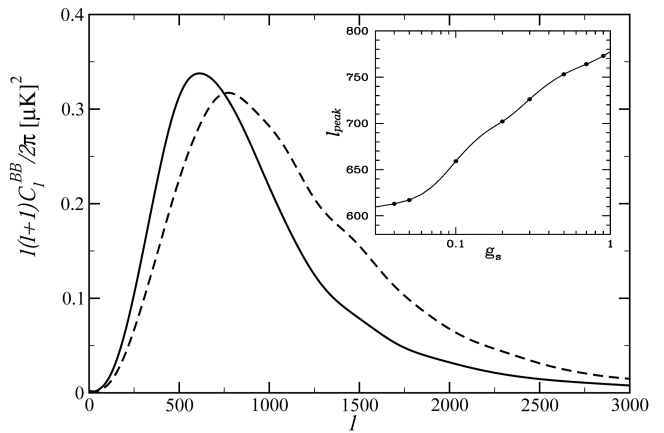}
\caption{\label{figBmode}The B-mode power spectra for $g_s=0.04$ (solid lines) and $g_s=0.9$ (dashed lines) normalised to $f_s=0.1$. The insert plot shows the position of the peak as a function of $g_s$.}
\end{figure}

Fig.~\ref{figBmode} shows the B-mode spectra for our two extreme values of the string coupling, namely $g_s=0.04$ and $g_s=0.9$.  Our main result is a significant shift in the peak towards low multipole number as the string coupling is reduced.  
This has a physical interpretation in terms of the network behaviour.  As mentioned above, the BB peak location depends on the correlation length and rms velocity of the string type dominating the power spectrum density $M_i$.  Our results show (Fig.~\ref{figNetwork}) that the velocity of the dominant string type only varies between $0.6$ and $0.7$, so we expect the peak position to be determined mostly by the dominant correlation length $\xi_i$.  As we saw, the power spectrum density is dominated by F strings for large $g_s$, but at lower $g_s$ there is a transition and the rare (larger $\xi_i$), heavy D strings become more important.  As we move to even smaller $g_s$, the D-string correlation length increases (Fig.~\ref{figNetwork}, middle panel) and the peak moves further down in $\ell$. 
Our result implies that an observation of the B-mode peak could help to break the degeneracy between $\mu_F$ and $g_s$ in the constraint imposed by the normalisation condition (\ref{fs}).  This degeneracy may also be broken \cite{cmbmulti} by combining CMB with pulsar timing constraints \cite{Jenet}.  

\begin{figure}[h]
\centering
\includegraphics[height=6cm, width=8cm]{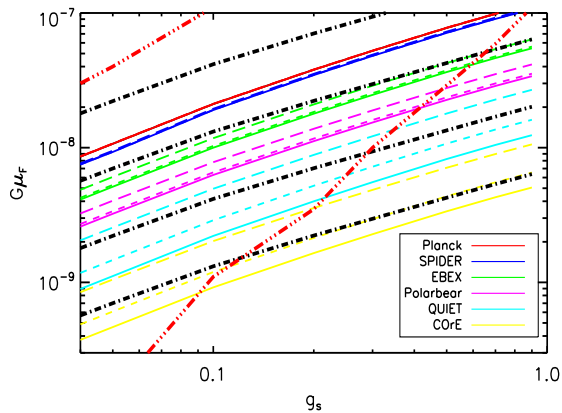}
\caption{\label{fig:sn-contours} The dot-dashed (black) curves show an $f_s$ of (from top to bottom) 0.1, 0.01, 0.001 and 0.0001 in the ($\mu_F$, $g_s$) parameter plane. Other curves show the S/N = 5 contours of an {\em overall} detection of the B-mode signal for various experiments.  The three curves for each experiment show the pessimistic (long-dashed lines), realistic (short-dashed lines) and optimistic (solid lines) scenarios for the removal of lensing B-modes, as discussed in the text. We also show the pulsar bounds based on data in \cite{Jenet} (dot-dot-dot-dashed lines) in the small (upper curve) and large (lower curve) string loop limits considered in~\cite{Battye:2010xz,cmbmulti}.}
\end{figure} 

\begin{figure}[h]
\centering
\includegraphics[scale= 0.5]{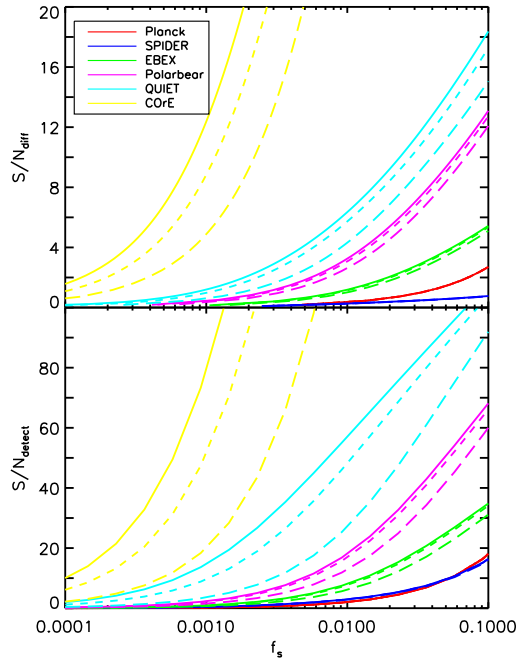}
\caption{\label{fig:sn_gs} The S/N of an overall detection (lower panel) as a function of $f_s$ for a reference model with $g_s = 0.1$. Labels are the same as in Fig.~\ref{fig:sn-contours}, with the sets of three curves (showing the optimistic, realistic and pessimistic lensing scenarios) from left to right.
The upper panel shows the S/N of the {\em difference} signal between two models with $g_s=0.04$ and $g_s=0.9$, with these both having the same overall detection S/N of the reference model at fixed $f_s$.}
\end{figure}
What are the chances of detection of the peak? We can estimate the ability of upcoming CMB polarisation experiments (using the relevant instrument sensitivities for each experiment) --- {\rm Planck}~\cite{lfi,hfi}, SPIDER~\cite{spider}, EBEX~\cite{ebex}, {\rm Polarbear}~\cite{polarbear} (using the parameters given in Ref.~\cite{Ma:20010yb}), QUIET~\cite{quiet} (using the phase 2 specification), as well as the proposed COrE~\cite{core} mission ---  to detect the position of the peak in the string sourced BB spectrum. We also refer the reader to a recent work in \cite{Mukherjee:2010ve} which studied the ability of the proposed CMBPol experiment \cite{baumann} to distinguish between B-modes sourced by textures, Abelian Higgs cosmic strings and gravitational waves sourced by inflation.
We follow the  procedure for obtaining the combined noise power spectrum for each experimental channel from Ref.~\cite{baumann}. In Fig.~\ref{fig:sn-contours} we show contours of constant signal-to-noise (S/N = 5) in the {\em overall} detection of the B-mode signal by each of the above experiments, depending on the values of the parameters $\mu_F$ and $g_s$.  Note that constraints on $f_s$ will be significantly tightened by measurements of TT, TE and EE from Planck and other experiments, but they cannot break the degeneracy between $\mu_F$ and $g_s$. For reference, we plot contours of $f_s=0.1$, $0.01$, $0.001$ and $0.0001$.  
Unfortunately the string B-mode signal can be obscured by those produced by gravitational lensing of the primary CMB. The sensitivity to strings will therefore be affected by the extent to which lensing can be removed. We account for this in three scenarios -- in the pessimistic case, we add the lensing component to the instrument noise power spectrum to obtain an effective noise, i.e. $N_{\ell}^{\rm eff} = N_{\ell} + C_{\ell}^{\rm lens}$. The lensing spectrum is computed using the CMB code CAMB for the best-fit WMAP7 $\Lambda$CDM model~\cite{Komatsu:2010fb}. In the second, we assume the added lensing signal can be reduced by a factor of 7 (although it may be more) in the white noise regime, as suggested by a quadratic estimator of the projected lensing potential~\cite{Seljak:2003pn}. 
Finally, we also show the most optimistic case where all the lensing is subtracted.  We note that contours of constant S/N closely match contours of constant $f_{\rm s}$. In the lower panel of Fig.~\ref{fig:sn_gs} we therefore present the S/N  as a function of $f_{\rm s}$ for a fixed reference model with $g_s = 0.1$. This can easily be interpreted in terms of  general $\mu_F$ and $g_s$ by referring to Fig.~\ref{fig:sn-contours}. Note that even without lensing subtraction, B-modes from strings can be detected by QUIET at high S/N even if strings contribute to TT just above the 0.1\% level. On the other hand, the string contribution would have to be close to the upper bound allowed by WMAP ($f_s \sim 0.05-0.1$)  to be detectable by {\rm Planck} and {\rm SPIDER}. The higher sensitivity of future experiments allows the possibility of distinguishing between models with different string couplings. We illustrate this by computing the {\em difference} signal between our two extreme values, $g_s=0.04$ (model A) and $g_s = 0.9$ (model B). To determine the S/N, the tension $\mu_{F}$ is adjusted for each of the two models, until they have the {\em same} overall detection ${\rm S/N}_{\rm detect}$ of B-modes as the reference model (the latter obtained from the lower panel of Fig.~\ref{fig:sn_gs}). The difference S/N is then computed via $\left( {\rm S/N}_{\rm diff} \right)^2 = \sum_{\ell} \left(C_{\ell}^{A} - C_{\ell}^{B}  \right)^2/ \left( \Delta C_{\ell}^{AB} \right)^2$\,, where the noise, including cosmic variance, is $ \Delta C_{\ell}^{AB} =\sqrt{2/(2\ell+1)} \left(C_{\ell}^A + C_{\ell}^B + N_{\ell}^{\rm eff} \right)$. We show this in the upper panel of Fig.~\ref{fig:sn_gs}. Since the difference signal is always more difficult to detect than an overall detection, ${\rm S/N}_{\rm diff} < {\rm S/N}_{\rm detect}$, with the degradation factor depending on the experiment and lensing removal.  We note that B-mode spectra are practically independent of other cosmological parameters, especially within the range currently allowed by WMAP,
because B-modes are sourced primarily by vector modes generated during the brief window of the last scattering surface.

We see that {\rm Planck} and {\rm SPIDER}  are unable to distinguish between the two models at high S/N, even if the string fraction is close to the WMAP limit. There is a bigger degradation in $S/N_{\rm diff}$ for {\rm SPIDER}, due to the lower angular resolution (the smallest SPIDER beam has an FWHM of $\sim 20'$, compared to $\sim 5'$ for {\rm Planck}), hence making  it more difficult to observe a shift in the B-mode peak. For EBEX the two models could be distinguished at low significance if the string fraction is close to the WMAP limit. For the {\rm Polarbear} and QUIET experiments, it is possible to distinguish between the two models with a $> 1 \%$ contribution to TT, depending on the delensing efficiency. COrE improves this discovery space by roughly an order of magnitude, and hence will be a powerful observatory for testing fundamental physics. We conclude with the optimistic thought that within the next generation of B-mode experiments we may well have the means to pin down the value of the string coupling constant, a direct observational test of string theory.

{\it Acknowledgements} ~We acknowledge discussions with Richard Battye, Clive Dickinson (QUIET), Samuel Leach (EBEX) and Sotirios Sanidas, and support from the CTC at U. of Cambridge (AA), the Royal Society (EJC), NSERC (AM and LP), U. of Manchester and U. of Nottingham (AP), and CNRS (DS and LP).

\end{document}